\documentclass[aps,preprint,superscriptaddress,showpacs]{revtex4}

\usepackage{amsmath}
\usepackage{bm}
\usepackage{graphicx}
\usepackage{amsfonts}
\usepackage{amssymb}

\begin{document}

\title{Dicke effect in a quantum wire with side-coupled quantum dots}

\author{P.\ A.\ Orellana}
\affiliation{Departamento de F\'{\i}sica, Universidad Cat\'{o}lica del Norte,
Casilla 1280, Antofagasta, Chile}

\author{F.\ Dom\'{\i}nguez-Adame}
\affiliation{GISC, Departamento de F\'{\i}sica de Materiales, Universidad
Complutense, E-28040 Madrid, Spain}

\author{E. Diez}
\affiliation{Departamento de F\'{\i}sica Fundamental, Universidad
de Salamanca, 37008 Salamanca, Spain}
\date{\today}

\begin{abstract}

A system of an array of side-coupled quantum-dots attached to a quantum wire is
studied theoretically. Transport through the quantum wire is investigated by
means of a noninteracting Anderson tunneling Hamiltonian. Analytical expressions
of the transmission probability and phase are given. The transmission
probability shows an energy spectrum with forbidden and allowed bands that
depends on the up-down asymmetry of the system. In up-down symmetry only the gap
survives, and in up-down asymmetry an allowed band is formed. We show that the
allowed band arises by the indirect coupling between the up and down quantum
dots. In addition, the band edges can be controlled by the degree of asymmetry
of the quantum dots. We discuss the analogy between this phenomenon with the
Dicke effect in optics.

\end{abstract}


\maketitle

\section{Introduction}

Quantum interference effects in quantum wires (QWs) are potentially useful in
nanotechnology since coupling to the continuum states shows an even-odd parity
effect in the conductance when the Fermi energy is localized at the center of
the energy band~\cite{dqd1,dqd2,qdn,yanson,qw,smit,oguri,zeng,kim}.
Consequently, a fine control of the electron transport can be achieved by
varying the external parameters of the QW.

In this context, we have recently considered  new quantum devices based on an
array of quantum dots (QDs)~\cite{pedro1}, a double QD~\cite{pss} and nanorings
\cite{pedro2} coupled to a QW. The attached device acts as scatterer for
electron transmission through the QW and allows to tune its transport
properties. It was found that the conductance at zero temperature through the QW
shows a complex behavior as a function of the Fermi energy: far from the center
of the band the conductance depends smoothly on the Fermi energy, while around
the center it develops an oscillating band with resonances and antiresonances
due to quantum interference in the ballistic channel. Moreover, the transmission
phase of the electron carries information complementary to the transmission
probability. This phase has been measured in QDs~\cite{yacoby,schuster} and
recently it was reported~\cite{sato} the experimental observation of the
Fano-Kondo antiresonance in a QW with a side-coupled QD. These experiments
proved that transport through the system has a coherent component.

In this work we report further progress along the lines indicated above. In
particular, we study theoretically transport properties of a set of side-coupled
double QDs attached to a perfect QW. We find an analytical expression for the
transmission probability and transmission phase. The transmission probability at
the center of the energy spectrum shows an energy spectrum with gap or allowed
band depending of the symmetry up-down, to be explained below. In a symmetry
up-down, an even-odd parity effect in the transmission phase at the center of
the band is demonstrated. Moreover, we show that an allowed band is formed in
the asymmetric case and that the width of this band can be controlled by
suitable gate voltages. This phenomenon is in analogy to  the Dicke effect in
quantum optics, that takes place in the spontaneous emission of two
closely-lying atoms radiating a photon into the same environment~\cite{dicke}.
In the electronic case, however, the decay rates (level broadening) are produced
by the indirect coupling of the up-down QDs, giving rise to a fast
(\emph{superradiant\/}) and a slow (\emph{subradiant\/}) mode. This close
analogy opens the way to exploit new electronic effects that usually arise in
atomic physics. In this regard, it has been shown that coupled QDs display the
electronic counterpart of Fano and Dicke effects that can be controled via a
magnetic flux~\cite{Orellana04}. Recently, Brandes reviewed the Dicke effect in
mesoscopic systems~\cite{brandes}.

\section{Model}

The system under consideration is shown in Fig.~\ref{fig0}. The QW is attached
to a $N$ side-coupled double QDs. The system, assumed in equilibrium, is modeled
by a noninteracting Anderson tunneling Hamiltonian~\cite{qdn} that can be
written as $H=H_{\text{QW}}+H_{\text{QD}}+H_{\text{QW-QD}}$, where
\begin{subequations}
\begin{equation}
H_{QW}=-v\sum_{i}\big(c_{i}^{\dagger}c_{i+1}^{}
+c_{i+1}^{\dagger}c_{i}^{}\big)\ ,
\end{equation}
describes the dynamics of the QW, $v$ being the hopping between neighbor sites
of the QW, and $c_{i}^{\dagger}$ ($c_{i}$) creates (annihilates) an electron
at site $i$.

\begin{figure}[ht]

\centerline{\includegraphics[width=60mm]{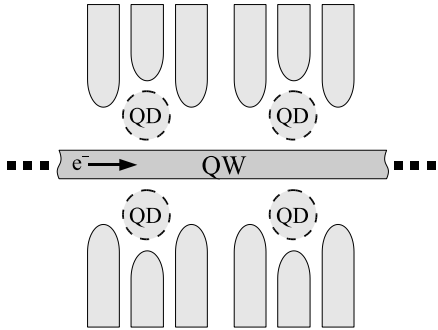}}
\caption{$N$ side-coupled double QDs attached to a perfect QW.}
\label{fig0}
\end{figure}

On the other side, $H_{D}$, given by
\begin{equation}
H_{D}=\sum_{j=1}^{N}\sum_{\alpha=u,d}
\varepsilon_{\alpha}d_{j\alpha}^{\dagger}d_{j\alpha}^{}
\end{equation}
is the Hamiltonian for the $N$ side-coupled double QDs, where $d_{j\alpha}$
($d_{j\alpha}^{\dagger}$) is the annihilation (creation) operator of an electron
in the QD ($j,\alpha)$,  $\varepsilon_{\alpha}$ is the corresponding single
energy level. Here the index $\alpha$ refers to the up ($u$) and down ($d$) QD,
attached at site $j$ of the QW. The coupling between the two subsystems (QW and
QDs) is described by the Hamiltonian
\begin{equation}
H_{D-W}=-V_{0}\sum_{j=1}^{N}\sum_{\alpha=u,d}\big(d_{j\alpha}^{\dagger}
c_{j}^{}+c_{j}^{\dagger}d_{j\alpha}^{}\big)\ ,
\end{equation}
where $V_{0}$ is the coupling between the QW and one of the QDs.

The Hamiltonian for the QW, $H_{QW}$, corresponds to a free-particle Hamiltonian
on a lattice with spacing unity and whose eigenfunctions are expressed as Bloch
solutions
\end{subequations}
\begin{equation}
\left\vert k\right\rangle =\sum_{j=-\infty}^{\infty}e^{ikj}
\left\vert j\right\rangle \ ,
\end{equation}
where $\left\vert k\right\rangle $ is the momentum eigenstate and $\left\vert
j\right\rangle $ is a Wannier state localized at site $j$. The dispersion
relation associated with these Bloch states reads
\begin{equation}
\omega=-2v\cos k\ .
\end{equation}
Consequently, the Hamiltonian supports an energy band from $-2v$ to $+2v$ and
the first Brillouin zone expands the interval $[-\pi,\pi]$. The stationary
states of the entire Hamiltonian $H$ can be written as
\begin{equation}
\left\vert \psi_{k}\right\rangle =\sum_{j=-\infty}^{\infty}a_{j}^{(k)}\left\vert
j\right\rangle + \sum_{j=1}^{N}\sum_{\alpha=u,d}
b_{j\alpha}^{(k)} \left\vert j,\alpha\right\rangle \ ,
\end{equation}
where the coefficient $a_{j}^{(k)}$ ($b_{j\alpha}^{(k)}$) is the probability
amplitude to find the electron at site $j$ of the QW [at the QD $(j,\alpha$)]
in the state $k$, that is, $a_{j}^{(k)}=\langle \,j|\psi_{k}\rangle$ and
$b_{j\alpha}^{(k)}=\langle \,j,\alpha|\psi_{k}\rangle$.

The amplitudes $a_{j}^{(k)}$ obey the following linear difference equation
\begin{subequations}
\begin{align}
\omega a_{j}^{(k)}  &  =v(a_{j+1}^{(k)}+a_{j-1}^{(k)})\ ,\quad j\leq 0
\mathrm{\ and\ }j>N\ ,\\
\omega a_{j}^{(k)}  &  =v(a_{j+1}^{(k)}+a_{j-1}^{(k)})-V_{0}\left(  b_{j,u}%
^{(k)}+b_{j,d}^{(k)}\right)  \ ,\ j=1,\ldots,N\ ,\\
\left(  \omega-\varepsilon_{\alpha}\right)  b_{j\alpha}^{(k)}  &  =-V_{0}%
a_{j}^{(k)}\ ,\ j=1,\ldots,N,\quad \alpha=u,d\ .
\label{eq-aj}
\end{align}
The amplitudes $b_{j,\alpha}^{(k)}$ can be expressed in terms of $a_{j}^{(k)}$
as follows
\end{subequations}
\begin{equation}
b_{j\alpha}^{(k)}=-\frac{V_{0}}{\omega-\varepsilon_{\alpha}}\,a_{j}^{(k)}\ ,
\label{bj-aj}%
\end{equation}
From Eq.~(\ref{bj-aj}) above, Eq.~(\ref{eq-aj}) becomes
\begin{equation}
(\omega-\tilde{\varepsilon})\,a_{j}^{(k)}=
v\left(a_{j+1}^{(k)}+a_{j-1}^{(k)}\right)\ ,
\quad j=1,\ldots,N\ ,
\label{newdif2}
\end{equation}
where the site energy $\tilde{\varepsilon}\equiv V_{0}^{2}/\left[ \left(
\omega-\varepsilon_{u}\right) +1/\left( \omega-\varepsilon_{d}\right) \right]$
depends on the electron energy $\omega$. Thus, the problem reduces to a linear
chain of a $N$ sites of effective energies $\tilde{\varepsilon}$. In order to
study the solutions of Eq.~(\ref{newdif2}), we assume that the electrons are
described by a plane wave incident from the far left with unity amplitude and a
reflection amplitude $r$, and at the far right by a transmission amplitude $t$.
That is,
\begin{align}
a_{j}^{(k)}  &  =e^{ikj}+re^{-ikj}\ , & j  &  <1\ ,\nonumber\\
a_{j}^{(k)}  &  =te^{ikj}\ , & j  &  >N\ .
\label{solut2}
\end{align}

The solution in the region $j=1,\ldots,N\ ,$ can be written as
\begin{widetext}
\begin{align}
a_{j}^{(k)}  &  =Ae^{iqj}+Be^{-iqj}\ , & \text{if }\left\vert (\omega
-\tilde{\varepsilon})/2v\right\vert  &  \leq1\ ,\quad q=-\cos^{-1}
\left[-(\omega-\tilde{\varepsilon})/2v\right]\ ,\nonumber \\
a_{j}^{(k)}  &  =Ce^{\kappa j}+De^{-\kappa j}\ , & \text{if }\left\vert
(\omega-\tilde{\varepsilon})/2v\right\vert  &  >1\ ,\quad \kappa=-\cosh^{-1}
\left[-(\omega-\tilde{\varepsilon})/2v\right] \ .
\label{solut}
\end{align}
\end{widetext}
Inserting~(\ref{solut2}) and~(\ref{solut}) into~(\ref{newdif2}), we get a
inhomogeneous system of linear equations for $A,B,C,D,t$ and $r$, leading to
the following result: If $|(\varepsilon-\tilde{\varepsilon})/2v|\leq1$,
\begin{subequations}
\begin{equation}
t=(-2ie^{ikN}/\Delta)\sin{k}\ ,
\end{equation}
\smallskip\noindent with $\Delta$ given by
\begin{equation}
\Delta=e^{-ik}\,\frac{\sin\left({N+1}\right){q}}{\sin q}-2\,\frac{\sin{Nq}}
{\sin q}+e^{ik}\,\frac{\sin\left({N-1}\right){q}}{\sin q}\ ,
\end{equation}
On the contrary, when $|(\varepsilon-\tilde{\varepsilon})/2v|> 1$
\begin{equation}
t=(2ie^{-ikN}/\Delta)\sin{k}\ ,
\end{equation}
\smallskip\noindent with $\Delta$ given by
\begin{equation}
\Delta=e^{-ik}\,\frac{\sinh\left(  {N+1}\right)  {\kappa}}{\sinh\kappa}
-2\,\frac{\sinh{N\kappa}}{\sinh\kappa}+e^{ik}\,\frac{\sinh\left(  {N-1}\right)
{\kappa}}{\sinh\kappa}\ ,
\end{equation}
\end{subequations}

The transmission probability is given by $T=\left\vert t\right\vert ^{2}$, and
it is related to the linear conductance at the Fermi energy  $\varepsilon_{F}$
by the one-channel Landauer formula at zero temperature,
$G=(2e^{2}/h)\,T(\omega=\varepsilon_{F})$~\cite{datta}. We also can obtain the
transmission phase as $\phi_{t}=
\tan^{-1}\left(\operatorname{Im}t/\operatorname{Re}t\right)$.

\section{Results}

To uncover the main features of the electron transport through the QW and
the effects of the attached QDs, we now consider several physical situations.
If $|(\omega-\tilde{\varepsilon})/2v|\leq1$, the transmission probability
and transmission phase reduces to
\begin{subequations}
\begin{align}
T &  =\frac{1}{\cos^{2}(Nq)+\big(\sin(Nq)\cot(k/2)/\sin q\big)^{2}%
}\ ,\label{g1}\\
\phi_{t} &  =\arctan\left\{  \frac{\alpha\cos{k\cos Nk+\beta}\sin k\sin
Nk}{\alpha\cos{k\sin Nk-}\beta\sin k\cos Nk}\right\} \ ,
\label{phase1}
\end{align}
with $\alpha   =\sin\left({N+1}\right){q}-2\sin{Nq}+\sin\left(
{N-1}\right){q}$ and  $\beta   =\sin\left(  {N+1}\right){q}-\sin\left(
{N-1}\right){q}$. We note that the transmission probability  oscillates as a
function of both $N$ and $q$. On the other hand, when
$|(\varepsilon-\tilde{\varepsilon})/2v|>1$ we get
\begin{align}
T  &  =\frac{1}{\cosh^{2}(N\kappa)+\big(\sinh(N\kappa)\cot(k/2)/\sinh
\kappa\big)^{2}}\ .
\label{g2}\\
\phi_{t}  &  =\arctan\left\{  \frac{\delta\cos{k\cos Nk+\eta}\sin k\sin
Nk}{\delta\cos{k\sin Nk-}\eta\sin k\cos Nk}\right\}  \ ,
\label{phase2}
\end{align}
\end{subequations}
where $\delta =\sinh\left(  {N+1}\right)  {\kappa}-2\sinh{N\kappa}+\sinh\left(
{N-1}\right)  {\kappa}$ and  $\eta =\sinh\left(  {N+1}\right)
{\kappa}-\sinh\left(  {N-1}\right) {\kappa}$.
Therefore, in this energy region $N$ tends exponentially to zero when $N$ is
large, namely $T\sim e^{-2N\kappa}$.

\begin{figure}[ht]
\centerline{\includegraphics[width=80mm,angle=-90]{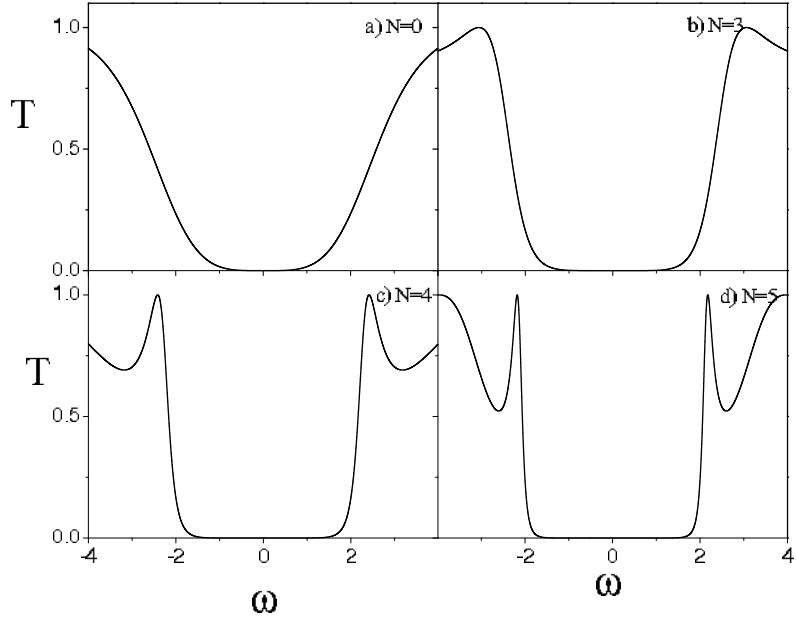}}
\caption{Transmission probability as a function of energy, in units of
$\gamma=V^{2}_{0}/2v$, for $\Delta V = 0$ and a)~$N=2$, b)~$N=3$, c)~$N=4$ and
d)~$N=5$.}
\label{fig2}
\end{figure}

To avoid the profusion of free parameters, for the sake of clarity we set the
energies of the up and down QDs as, $\varepsilon_{u}=\Delta V$ and
$\varepsilon_{d}=-\Delta V$ hereafter. We first consider the case $\Delta V=0$.
In this case the transmission exhibits a forbidden band (gap). Figure~\ref{fig2}
shows the transmission probability versus $\omega$ for different values of $N$.
It is apparent that $T$ tends to zero within a range $[-2\gamma,2\gamma]$, with
$\gamma=V_{0}^{2}/2v$, and the system shows a gap of width $4\gamma$.

Inside the energy region $[-2\gamma,2\gamma]$ the conductance does not present
any feature that depends on $N$. However, at the center of the band ($\omega=0$,
$k=\pi/2$, $\kappa\rightarrow\infty$) the transmission phase takes on the
following values, depending on the parity of $N$: When $N$ is odd, $\phi_{t} \to
\pm \pi/2$ for $\omega \to \mp 0$ , while for $N$ even, $\phi_{t} =0$ for
$\omega = 0$. We can understand this result as follows. Each time that the
electron passes near a double QDs, it undergoes a phase change equal to $\pi/2$
due to the destructive interference between the discrete levels in the double QD
and the continuum states of the QW.

\begin{figure}[ht]
\centerline{\includegraphics[width=80mm,angle=-90]{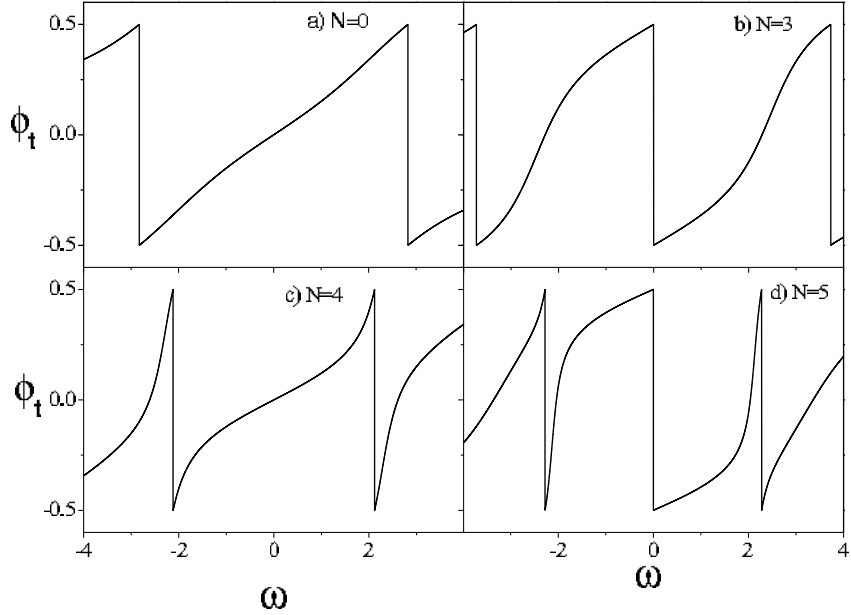}}
\caption{Transmission phase versus energy, in units of $\gamma=V^{2}_{0}/2v$,
for $\Delta V = 0$ and a)~$N=2$, b)~$N=3$, c)~$N=4$ and d)~$N=5$.}
\label{fig3}
\end{figure}

Consider now the situation with $\Delta V\neq 0$. Figure~(\ref{fig3}) shows the
transmission probability for different values of $\Delta V$, for $N=4$ and
$N=5$. We note that an allowed band develops at the center of the gap. It is
straightforward to show that for $\Delta V\ll\gamma$ the width of this allowed
band is $\Delta V^{2}/2\gamma$. Moreover the transmission  probability becomes
always unity at the center of the allowed band, independently of the value of
$N.$

\begin{figure}[ht]
\centerline{\includegraphics[width=80mm,angle=-90]{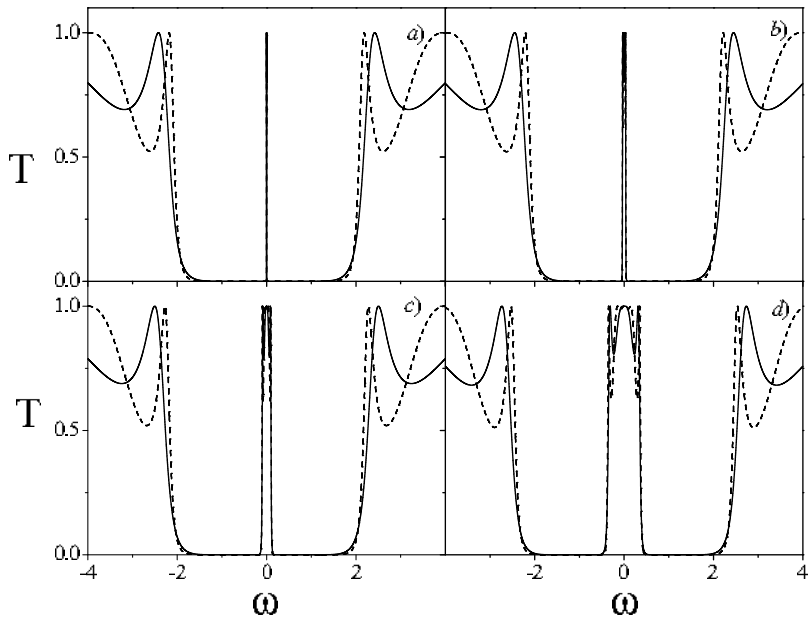}}
\caption{Transmission probability versus energy, in units of $\gamma$ for
$N=4$(solid line) and $N=5$ (dashed line) for a)~$\Delta V = 0.1 \gamma$,
b)~$\Delta V = 0.3 \gamma$, c)~$\Delta V = 0.5 \gamma$ and  d)~$\Delta V = 1.0
\gamma$.}
\label{fig4}
\end{figure}

\begin{figure}[ht]
\centerline{\includegraphics[width=80mm,angle=-90]{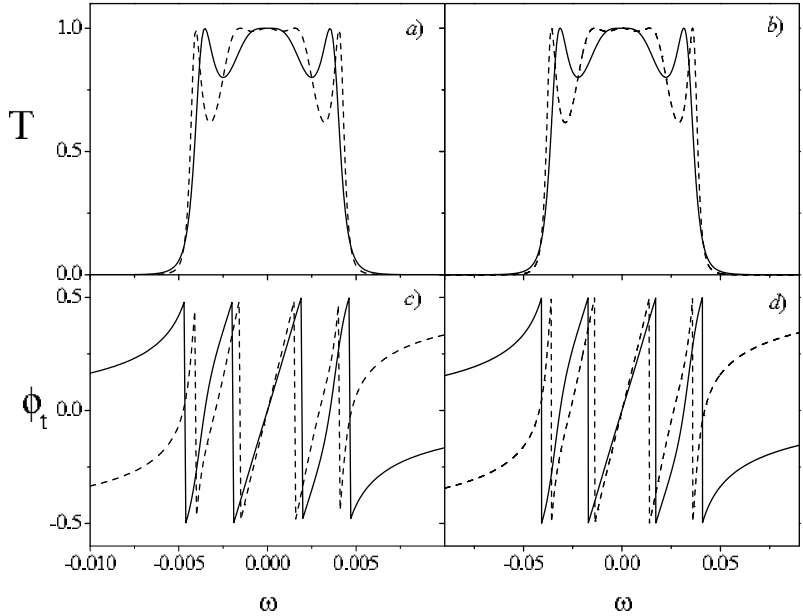}}
\caption{Enlarged views of the transmission probability and transmission phase
versus energy, in units of $\gamma$, for $N=4$(solid line) and $N=5$ (dashed
line) for $\Delta V = 0.1 \gamma$ (left panels) and  $\Delta V = 0.3 \gamma$
(right panels).}
\label{fig5}
\end{figure}

Figure~\ref{fig4} displays the transmission probability for several values of
$\Delta V$ when $N=4$ (solid line) and $N=5$ (dashed line). We note that the
shape of the allowed band is independent of the value of $\Delta V$.
Additionally,  Fig.~\ref{fig5} shows the transmission phase for a similar set of
parameters in the region of the allowed band. We note a series of
discontinuities of the transmission phase at some values of the energy $\omega$.
It is worth to note that the drops in the transmission phase do not imply
vanishing transmission probability. At the energies of the drops of the
transmission phase, the real part of the transmission amplitude vanishes and at
the same time its imaginary part changes its sign. Additionally from
Eq.(\ref{phase1}), we obtain that the transmission phase is zero (or $m\pi$, $m$
integer) at the center of the allowed band, independent of the value of $N$ and
$\Delta V$. At the same time, the transmission probability becomes unity.

This phenomenon resembles the Dicke effect in optics, which takes place in the
spontaneous emission of a pair of atoms radiating a photon with a wave length
much larger than the separation between them~\cite{dicke}. The luminescence
spectrum is characterized by a narrow and a broad peak, associated with long and
short-lived states, respectively. The former state, weakly coupled to the
electromagnetic field, is called \emph{subradiant}, and the latter, strongly
coupled, \emph{superradiant\/} state. In the present case this effect is due to
the indirect coupling between up-down QDs through the QW. The states strongly
coupled to the QW yield a forbidden band with width $4\gamma$ and the states
weakly coupled to the QW give an allowed Dicke band with width $\Delta
V^{2}/2\gamma$.

\section{Summary}

In this work we studied the transmission probability and transmission phase
through a QW with an array of side-coupled double QDs. We found that the
transmission probability displays a gap or a band at the center of the  energy
spectrum, due to destructive and constructive interference in the ballistic
channel, respectively. For an array with symmetry up-down ($\Delta V=0$) only a
gap develops at the center of the band. For $\Delta V\neq0$ an allowed band
arises at the center of the energy spectrum independently of the value of
$\Delta V$. This phenomenon is in analogy to the Dicke effect in optics. The set
of side coupled double QDs seems a suitable system to study the Dicke effect in
experiments on quantum transport. This effect could be used to develop
nanodevices where an extremely fine control of electron transport could be
achieved just by varying the gate potential of the QDs.

\begin{acknowledgments}

P.\ A.\ O.\ would like to thank financial support from Millennium Science
Nucleus Condensed Matter Physics and FONDECYT under grants Nos. 1020269 and
7020269. Work in Madrid was supported by MCyT (MAT2003-01533). E. Diez
acknowledge support by Junta de Castilla y Le\'{o}n (SA007B05) and MEC (Ram\'on
y Cajal and FIS2005-01375). Additionally, P.\ A.\ O.\ would like to thank the
hospitality of Departamento de F\'{\i}sica de Materiales (UCM) and Departamento
de F\'{\i}sica Fundamental (USAL) during his visits.

\end{acknowledgments}

\end{document}